# COOL DWARFS IN WIDE MULTIPLE SYSTEMS

## PAPER 3: TWO COMMON-PROPER-MOTION, LATE-TYPE STARS SEPARATED BY OVER 11 ARCMIN

*By José A. Caballero*
*Centro de Astrobiología (CSIC-INTA), Madrid,*
*J. Genebriera*
*Observatorio de Tacande, La Palma,*
*F. X. Miret, T. Tobal, J. Cairol*
*Observatori Astronòmic del Garraf, Barcelona*

LP 209–28 and LP 209–27 have similar proper motions as tabulated by several catalogues. Using seven astrometric epochs spanning 59 years, we confirm a common tangential velocity by measuring a constant angular separation of $\rho$ = 666.62±0.09 arcsec. Accurate SDSS and 2MASS photometry indicates that they are normal dwarfs of approximate spectral types K7 V and M3 V. However, from their apparent magnitudes, both LP 209–28 and LP 209–27 are located at 200–250 pc, from where one can deduce an astonishing projected physical separation of 0.6–0.8 pc. The system Koenigstuhl 6 AB represents another world record among the least-bound systems with low-mass star components.

*Koenigstuhl 6 AB: an extremely fragile system of two late-K/early-M dwarfs*

After the discovery of Koenigstuhl 4 AB (two mid-M dwarfs in a loosely-bound common-proper-motion pair)[1] and Koenigstuhl 5 AB-C (a distant M8.5V companion to a binary solar-type star)[2], now we present the third paper of the series 'Cool dwarfs in wide multiple systems', with the identification of two red Luyten stars sharing the same proper motion and being separated by over 11 arcmin. We call it Koenigstuhl 6 AB (abridged: Kö6AB) and it is one of the most prominent outcomes of a professional-amateur virtual-observatory search for companions to Luyten stars[3].

LP 209–28 (Kö6A) and LP 209–27 (Kö6B) are two poorly-known red stars for which Luyten[4,5] estimated spectral types M from $B_J$ and $R_F$ photographic magnitudes. Available information in the literature on the stars is optical and near-infrared magnitudes and proper motions only[6,7]. None of them are tabulated by the *Washington Double Star Catalogue*[8]. We compile in Table I their basic data, including coordinates and proper motions from the PPMXL Catalogue of Positions and Proper Motions[9], optical photometry from the SDSS Sloan Digital Sky Survey[10] and near-infrared photometry from the 2MASS Two-Mass All-Sky Survey[11]. The PPMXL proper motion, of over 200 mas a$^{-1}$, is identical, within uncertainties, to other proper motions tabulated in the literature[12,13].

We measured angular separations, $\rho$, and position angles, $\theta$, for seven astrometric epochs between 1953 and 2012 (Table II). The six first epochs corresponded to the SuperCOSMOS digitizations of the Palomar Observatory Sky Survey[14], 2MASS and SDSS (marked with an asterisk), which was actually the mean of two SDSS observations separated by 24 days. The seventh and last astrometric epoch was taken by us on 2012 February 15 using a dual CCD camera SBIG ST-8XME at the 0.4 m Telescopio del Observatorio de Tacande in La Palma, Spain (Minor Planet Center code J22, http://www.astropalma.com/), as described by Caballero *et al.*[15]. The camera provided an 18.1 × 12.1 arcmin$^2$ field of view with a pixel size of 0.71 arcsec and first-order adaptive optics (tip-tilt). We took three consecutive exposures of 300 s in the SDSS *r'* band when the pair culminated (airmass = 1.02) and astrometrically calibrated the stacked image with background 2MASS sources of null proper motion. As summarized in the last row of Table II, both $\rho$ and $\theta$ between the two stars kept constant with very small standard deviations, of only 0.09 arcsec and 0.010 deg, respectively. Thus, Kö6A and Kö6B share the same high proper motion.

From the SDSS and 2MASS photometry of the two stars in Table I and a simple inspection of a very large sample of cool dwarfs with the same photometry and low-resolution spectroscopy[16], we estimated spectral types K7: V and M3: V for Kö6A and Kö6B, respectively. Uncertainties are less than two spectral sub-types.

Instead of assuming a spectral-type–absolute-magnitude relation, in Table III we derived the basic astrophysical parameters of Kö6AB as functions of distance. Masses and effective distances were obtained from the *J*-band absolute magnitude and NextGen98 theoretical models of the Lyon group[17]. Derived effective temperatures were too cool and too warm for K7 V and M3 V stars at heliocentric distances *d* = 150 and 300 pc, respectively. Only at *d* = 200–250 pc, spectral types estimated from SDSS-2MASS colours and effective temperatures from the *J*-band absolute magnitudes matched within reasonable uncertainties. Next, we assumed that the pair Kö6AB is located at a distance in the 200–250 pc interval. A more precise determination of the heliocentric distance shall be obtained by an accurate spectral type determination together with the use of a reliable spectral-type–absolute-magnitude relation[18,19,20] or, far better, from a direct parallax measurement, in particular by the ESA *Gaia* mission.

At the assumed distance interval, the angular separation translates into a projected physical separation of 0.6–0.8 pc. While wider systems have been proposed recently[21,22,23], the total mass of Kö6AB is lower than them (*i.e.*, the Zubenelgenubi multiple system[20] has around 6.7 $M_{sol}$, in comparison with the approximately 0.88-1.04 $M_{sol}$ of Kö6AB). This makes our system have a very low absolute magnitude of binding energy $U^*_g$ (which is proportional to the product of component masses and inversely proportional to the *projected* physical separation[21]), at the boundary between very wide binaries and stellar kinematic groups, which seems to lay at approximately $-10^{33}$ J (see again works by Caballero[21,22]). The scarcity of data on Kö6A and Kö6B prevents from inferring any age, which usually comes from X-ray, spectroscopic (H$\alpha$, Li I) or kinematic (*U, V, W*) data. Few young moving associations are known further than 100 pc (*e.g.*, Torres *et al.*[24] and references therein). A radial-velocity determination for both stars will erase any doubt on its common velocity and help establishing their original stellar population in the Galaxy. In other words, we do not know yet whether Kö6AB is an old binary in the disc or halo caught in disintegration or a relatively young pair of stars originated from the same parental cloud, from where they were ejected in the same direction, but, in any case, Kö6AB represents another milestone in the search for the least bound binary systems.

Table I
*Basic data of Koenigstuhl 6 and B*

| *Datum* | *A* | *B* | *Origin* |
|---|---|---|---|
| Name | LP 209–28 | LP 209–27 | Luyten |
| NLTT | 19878 | 19858 | Luyten |
| $\alpha$ (J2000) | 08 37 04.69 | 08 36 37.43 | PPMXL |
| $\delta$ (J2000) | +39 07 59.8 | +38 58 13.6 | PPMXL |
| $\mu_\alpha \cos\delta$ [mas a$^{-1}$] | –126.2±4.0 | –123.9±4.0 | PPMXL |
| $\mu_\delta$ [mas a$^{-1}$] | –194.4±4.0 | –182.0±4.0 | PPMXL |
| $u$ [mag] | 19.152±0.028 | 21.111±0.111 | SDSS |
| $g$ [mag] | 16.615±0.004 | 18.841±0.009 | SDSS |
| $r$ [mag] | 15.134±0.004 | 17.456±0.006 | SDSS |
| $i$ [mag] | 14.602±0.009 | 16.787±0.005 | SDSS |
| $z$ [mag] | 13.981±0.004 | 15.339±0.006 | SDSS |
| $J$ [mag] | 12.816±0.022 | 14.009±0.026 | 2MASS |
| $H$ [mag] | 12.227±0.022 | 13.478±0.029 | 2MASS |
| $K_s$ [mag] | 12.054±0.020 | 13.246±0.031 | 2MASS |
| Phot. Sp. Type | K7: V | M3: V | This work |

Table II
*Astrometric observations of the Koenigstuhl 6 AB system*

| Epoch | $\rho$ [arcsec] | $\theta$ [deg] | Origin |
|---|---|---|---|
| 1953 Mar 16 | 666.72 | 208.417 | POSSI Red |
| 1989 Mar 09 | 666.62 | 208.410 | POSSII Red |
| 1991 Feb 04 | 666.66 | 208.423 | POSSII Infrared |
| 1994 Jan 11 | 666.62 | 208.413 | POSSII Blue |
| 1998 Apr 04 | 666.51 | 208.416 | 2MASS |
| 2001 Dec 03* | 666.71 | 208.430 | SDSS |
| 2012 Feb 15 | 666.48 | 208.395 | This work |
| *Average* | 666.62±0.09 | 208.415±0.011 | This work |

Table III
*Basic data of the Koenigstuhl 6 AB system as functions of distance*

| | A | B | A | B | A | B | A | B |
|---|---|---|---|---|---|---|---|---|
| $d$ [pc] | 150 | | 200 | | 250 | | 300 | |
| $s$ [AU] | 100000 | | 133300 | | 166700 | | 200000 | |
| $m-M$ [mag] | 5.880 | | 6.505 | | 6.990 | | 7.386 | |
| $M_J$ [mag] | 6.936 | 8.128 | 6.311 | 7.504 | 5.826 | 7.019 | 5.430 | 6.623 |
| $\mathcal{M}$ [M$_{sol}$] | 0.44: | 0.25: | 0.54: | 0.34: | 0.61: | 0.43: | 0.68: | 0.49: |
| $T_{eff}$ [K] | 3570: | 3380: | 3730: | 3470: | 3940: | 3560: | 4130: | 3640: |
| $\mathcal{M}_{total}$ [M$_{sol}$] | 0.69: | | 0.88: | | 1.04: | | 1.17: | |
| $-U^*_g$ [10$^{33}$ J] | −1.9 | | −2.4 | | −2.8 | | −2.9 | |

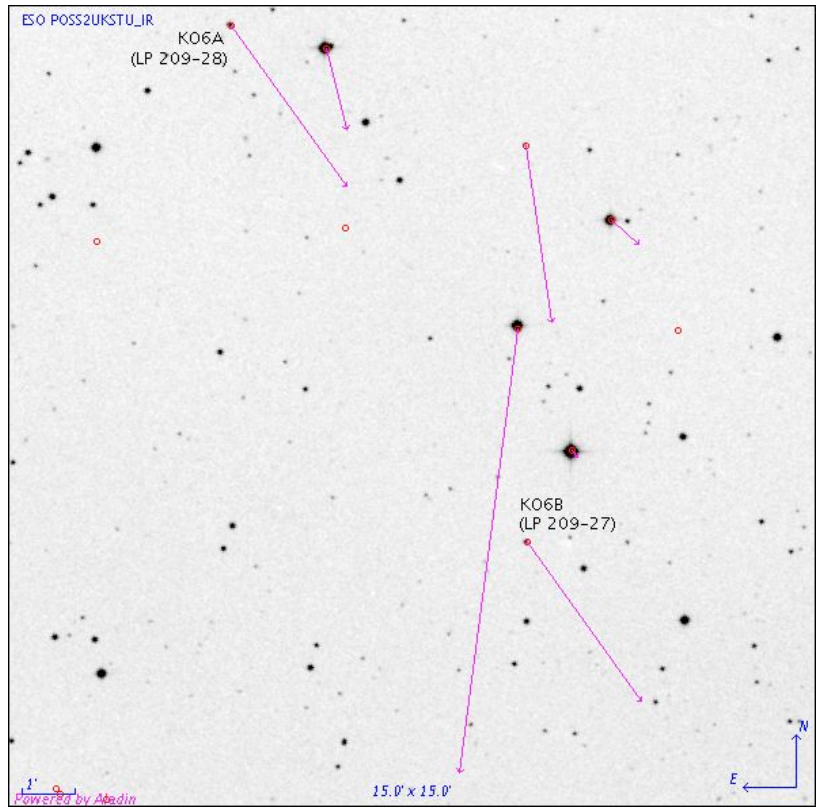

FIG. 1
Colour-inverted UKST $I_N$-band image provided by *The Digitized Sky Survey* from the European Southern Observatory at Garching, constructed with *Aladin* showing Koenigstuhl 6 A (LP 209–28; top left) and B (LP 209–27; top right), which are labelled. The long arrows indicate the proper motions of stars as tabulated by *Simbad*. Field of view is 15 arcmin × 15 arcmin; the orientation is indicated on the image. The star with the highest proper motion, roughly in the middle of the field, is the foreground M:-type star G 115–11.